\documentclass[aps,prd,twocolumn,floatfix,nofootinbib,superscriptaddress,showkeys]{revtex4}
\usepackage{dcolumn}
\usepackage{amsmath,amsfonts,extarrows,amssymb,mathrsfs,times,bm}
\usepackage{extarrows}
\usepackage{graphicx}
\usepackage{float}
\usepackage{graphicx,epstopdf}
\usepackage{dcolumn}
\usepackage{exscale}
\usepackage{relsize}
\usepackage{mathtools}
\usepackage{mhchem}
\usepackage[colorlinks=true,breaklinks=true,linkcolor=blue,citecolor=blue,urlcolor=blue]{hyperref}
\newcommand{\nn}{\nonumber}

\begin{document}
\title{Kerr-Newman Jacobi geometry and the deflection of charged massive particles}
\author{Zonghai Li}
\affiliation{Center for Astrophysics, School of Physics and Technology, Wuhan University, Wuhan 430072, China}
\date{\today}
\author{Junji Jia}
\email[Corresponding author:~]{junjijia@whu.edu.cn}
\affiliation{Center for Astrophysics \& MOE Key Laboratory of Artificial Micro- and Nano-structures, School of Physics and Technology, Wuhan University, Wuhan, 430072, China}

\begin{abstract}
In this paper, we investigate the deflection of a charged particle moving in the equatorial plane of Kerr-Newman spacetime, focusing on weak field limit. To this end, we use the Jacobi geometry, which can be described in three equivalent forms, namely Randers-Finsler metric, Zermelo navigation problem, and $(n+1)$-dimensional stationtary spacetime picture. Based on Randers data and Gauss-Bonnet theorem, we utilize osculating Riemannian manifold method and the generalized Jacobi metric method to study the deflection angle, respectively.
In the $(n+1)$-dimensional spacetime picture, the motion of charged particle follows the null geodesic, and thus we use the standard geodesic method to calculate the deflection angle. Three methods lead to the same second-order deflection angle, which is obtained for the first time. The result shows that the black hole spin $a$ affects the deflection of charged particles both gravitationally and magnetically at the leading order (order $\mathcal{O}([M]^2/b^2)$). When $qQ/E<2M$, $a$ will decrease (or increase) the deflection of prograde (or retrograde) charged signal. If $qQ/E> 2M$, the opposite happens, and the ray is divergently deflected by the lens. We also showed that the effect of the magnetic charge of the dyonic Kerr-Newman black hole on the deflection angle is independent of the particle's charge.
\end{abstract}

\keywords{Deflection angle, Charged particle, Kerr-Newman spacetime, Jacobi metric, Randers-Finsler metric, Zermelo navigation problem, Gauss-Bonnet theorem}
\maketitle

\section{Introduction}
The deflection of light caused by gravitational field is a prediction of general relativity, and it was observed by Eddington's team in 1919~\cite{DED1920,Will2015}. In addition to testing theories of gravity, the deflection effect can also be used to distinguish between a wormhole, naked singularity and black hole~\cite{Tsukamoto2012,Tsukamoto2013,Tsukamoto2018,JusufiAmir} and to study the thermodynamics of AdS black holes~\cite{Belhaj2021}. Moreover, based on the deflection of light, gravitational lensing has become a powerful tool to measure the
mass of galaxies and clusters~\cite{Hoekstra2013,Brouwer2018,Bellagamba2019}, and to search for dark matter and dark energy~\cite{Vanderveld2012,cao2012,zhanghe2017,Huterer2018,SC2019}. The usual way to study the gravitational deflection angle is to calculate the null/timelike geodesic in 4-spacetime, that is, the standard geodesic method~\cite{Weinberg}. Recently, a widely popular optical metric method (OMM) using the differential geometrical formalism in 3-space defined by the optical metric, has been proposed by Gibbons and Werner~\cite{GW2008,Werner2020}.

Optical geometry (also called optical reference geometry or Fermat geometry) was first introduced by Weyl in 1917~\cite{Weyl}. For an $(n+1)$-dimensional static spacetime with metric
\begin{eqnarray}
ds^2=g_{tt}dt^2+g_{ij}dx^idx^j,~~i,j=1,2,\cdots,n
\end{eqnarray}
the optical metric reads
\begin{eqnarray}
dt^2=-\frac{g_{ij}}{g_{tt}}dx^idx^j.
\end{eqnarray}
According to Fermat's principle, the motion of light in this $(n+1)$-dimensional spacetime is governed by the geodesic of an $n$-dimensional optical space.
The main point of the OMM in Ref.~\cite{GW2008} was to link the geometric properties of optical metrics with the gravitational lensing, which is achieved through the application of the Gauss-Bonnet (GB) theorem. As a result, the weak gravitational deflection angle of light can be obtained by integrating the Gaussian curvature of the optical metric. This method shows that the gravitational lensing can be viewed as a global effect.
In addition, the topological effects of light ray was studied by optical geometry and GB theorem by Gibbons, Werner and their collaborators~\cite{Gibbons and Warnick,Petters and Werner}. The OMM pioneered by Gibbons and Werner has evolved into an active direction of research. Works about spacetimes with different symmetry and asymptoticity, and signals of different types have been carried out. For example, the deflection of light via optical metric and GB theorem has been explored in different static spacetimes such as Ellis wormhole and Janis-Newman-Winnicour wormhole spacetimes~\cite{Jusufi:wormhole}, charged wormhole spacetime in Einstein-Maxwell-dilaton theory~\cite{Jusufi:emd}, as well as some asymptotically non-flat spacetimes~\cite{Ali:wormhole,Ali:strings,Ali:BML}. Other contribution include deflection of light in a plasma medium~\cite{Crisnejo;Gallo}, the influence of Brane-Dicke coupling parameter, dilaton field and nonlinear electrodynamic on lensing of light~\cite{Javed1,Javed2,Javed3}, and so on. Furthermore, Ishihara et al. used optical metric and GB theorem to study the finite distance deflection of light in static gravitational field both in the weak and strong deflection limits~\cite{ISOA2016,IOA2017}. In this case, the receiver and source are assumed to be at finite distance from a gravitational lens. Another study on the finite distance deflection of light can be found in the work of Arakida~\cite{Arakida2018}. Furthermore, the initial OMM was also extended to stationary spacetimes because of their relevance in astrophysics.

In stationary spacetime whose optical geometry is defined by a Randers type Finsler metric~\cite{Randers1941} however, we encounter the difficulty of an intrinsically Finslerian description of the GB theorem. To solve this problem, Werner~\cite{Werner2012} constructed an osculating Riemannian manifold of the Randers-Finsler manifold through the Naz{\i}m's method~\cite{Nazim1936}, which can be used to study the propagation of light. Werner's method has been used to different stationary fields, for instance, rotating wormhole and rotating regular black holes~\cite{Jusufi&Ali:Teo,Jusufi:RB}, as well as asymptotically non-flat stationary fields such as rotating cosmic string and rotating global monopole~\cite{{Jusufi:string17,Jusufi&Ali:string,Jusufi:monopole}}. The other technique that can use GB theorem to calculate the deflection of light in a stationary spacetime is the so-called generalized OMM, established by Ono and his collaborators~\cite{OIA2017,OIA2018,OIA2019}. By Fermat's principle, one can assume that the light ray moves in a Riemannian space and is affected by a one-form. As a result, the motion of light no longer follows the geodesic in the Riemannian space, so the influence of the geodesic curvature on the deflection angle needs to be considered. The generalized OMM was popular in recent studies~\cite{Ali:OIA1,Ali:OIA2,Kumar2019,Haroon2019}.

In general, through optical geometry and GB theorem, the weak gravitational deflection problem of light can be solved elegantly. From both theoretical and experimental point of view, in addition to photons, people are also interested in the deflection and lensing of massive particles. A natural consideration is to extend this geometric method to the particle case. To this end, we need to use the Jacobi metric formalism. Base on principle of least action of Maupertuis, the trajectories of a given mechanical system of constant total energy, are geodesic of the Jacobi metric. Jacobi metric is one of the main tools of geometric dynamics and has been used to study various mechanical problems under the framework of Newton's theory~\cite{Pin1975,Szydlowski1996,Awrejcewicz}. Gibbons first established the Jacobi metric for a neutral massive particle moving in a static spacetime~\cite{Gibbons2016}. Chanda and his collaborators subsequently extended this work to stationary spacetime~\cite{Chanda2019}. As mentioned before, null geodesics in an $(n+1)$-dimensional spacetime corresponds to geodesics in the corresponding $n$-dimensional optical space. Similarly, timelike rays in an $(n+1)$-dimensional spacetime corresponds to geodesics in $n$-dimensional Jacobi space. The same principle holds in the presence of electromagnetic field~\cite{Das2017,Chanda2019b}. With Jacobi metric and GB theorem, Li and his collaborators studied the deflection of massive particle in both static and stationary spacetimes~\cite{LHZ2020,Lijia2020,LiA2020,LizhouRas,lidujia}. Among other things, Werner's method and the generalized OMM have been extended in these studies. In addition, Crisnejo, Jusufi and their collaborators used GB theorem to calculate the deflection angle of massive particle by establishing the correspondence between light ray in the medium and the particles ray in a spacetime~\cite{Crisnejo;Gallo,Crisnejodengren,Jusufimassive,CrisnejodengrenJusufi,Jusufi:cmp}.

In~\cite{lidujia}, we considered the deflection of charged particles in a charged static spacetime using the GB theorem. In this article we will further extend the study of charged particle deflection to stationary spacetime. The Jacobi geometry of a charged particle in charged stationary spacetime is also defined by a Randers-Finsler metric. Therefore, we can use the osculating Riemannian manifold method (ORMM) and the generalized Jacobi metric method (GJMM) to calculate the deflection angle of charged particle. Mathematically, Randers data and Zermelo data are equivalent. The solution of the Zermelo navigation problem on Riemannian manifold is a Randers metric. Conversely, any Randers metric corresponds to a Zermelo navigation problem \cite{Gibbons2009}. Therefore we will also state the Zermelo data equivalent to the Randers metric. There is also a third equivalent expression, pointed out by Gibbons and collaborators,
saying that the geodesic flow in an $n$-dimensional Randers space can be regarded as the null geodesic flow in the corresponding $(n+1)$-dimensional stationary spacetime ($(n+1)$DSS) \cite{Gibbons2009}. From this viewpoint, one can treat charged particles as photons in the $(n+1)$DSS, and then calculate its deflection angle via null geodesic. The most typical charged and rotating solution is the Kerr-Newman (KN) black hole~\cite{newman1,newman2}, characterized by its mass, spin angular momentum and charge. In literature, the lensing of KN spacetime has been widely discussed such as equatorial light ray~\cite{Chakraborty&Sen,LiZhou2020,Hsiao2020}, any arbitrarily incident directions light ray~\cite{Jiang&Lin}, and equatorial neutral massive particle~\cite{He&lin2016,He&lin20162666,He&lin2017,jiaepjc2020,kejia}. What is worth mentioning here is Jusufi's work on the deflection of charged particles in KN spacetime~\cite{Jusufi:cmp}. Jusufi used a Riemannian optical metric and GB theorem. However, Jusufi did not study the unique effect of black hole rotation on charged particles. In the present work we shall investigate this interesting question by calculating the deflection angle of charged particle in the equatorial plane of a KN black hole, using Werner's ORMM, the GJMM, and the standard geodesic method, respectively.

This paper is organized as follows. In Sec.~\ref{jacobimetric}, we shall first review the Jacobi-Randers metric for a charged particles in general stationary  spacetime. Then, we will discuss its other two equivalent descriptions, namely Zermelo navigation problem and the $(n+1)$DSS picture. Finally, we introduce the GB theorem and apply it to the lensing geometry to obtain a general formula for calculating the deflection angle. In Sec.~\ref{KNjacobi} we will derive for the KN spacetime its Jacobi geometry described by Randers
metric, Zermelo navigation problem, and the $(n+1)$DSS data, respectively. In Sec.~\ref{anglechen3}, we calculate the weak deflection angle of charged particles via geodesic method, Werner's ORMM, and the GJMM, respectively. Finally, we end our paper with a short conclusion in Sec.~\ref{conclusion}. Throughout this paper, we use the natural unit $G = c = 1/(4\pi\epsilon)=1$ and the spacetime signature $(+,-,-,-)$.

\section{Jacobi metric of charged particles in stationary spacetime and Gauss-Bonnet theorem}\label{jacobimetric}
In this section, we shall first review the Jacobi metric for a charged massive particle in a stationary spacetime according to Chanda~\cite{Chanda2019b}. Then, learning from Gibbons~et al.~\cite{Gibbons2009} we discuss the equivalent description of the Jacobi metric in Zermelo navigation problem and $(n+1)$DSS picture. Finally, we will introduce the GB theorem for curved surfaces and applied it to lensing geometry.

\subsection{Jacobi-Randers metric}
Let us begin by written the line element of a stationary spacetime
\begin{eqnarray}
d{s}^2={g}_{tt}(x)dt^2+2g_{ti}(x)dtdx^i+{g}_{ij}(x)dx^idx^j,
\end{eqnarray}
the Lagrangian of a charged particle of mass $m$, charge $q$ and energy $E$ can be written as
\begin{eqnarray}
\mathcal{L}&=&-m\sqrt{g_{\alpha\beta}\dot{x}^\alpha \dot{x}^\beta}+qA_{\alpha}\dot{x}^{\alpha},
\end{eqnarray}
where dot means derivative with respect to $t$ and $A_\alpha$ is the electromagnetic gauge potential. The momentum conjugate to $t$ and $x^i$ are respectively
\begin{eqnarray}
\label{momentum}
&&p_t=\frac{\partial \mathcal{L}}{\partial \dot{t}}=-\frac{m\left(g_{tt}\dot{t}+g_{ti}\dot{x}^i\right)}{\sqrt{g_{\alpha\beta}\dot{x}^\alpha \dot{x}^\beta}}+qA_t=-E,\\
\label{momentumli2}
&&p_{i}=\frac{\partial \mathcal{L}}{\partial \dot{x}^i}=-\frac{m \left( {g}_{ij}\dot{x}^j+g_{ti}\dot{t}\right)}{\sqrt{g_{\alpha\beta}\dot{x}^\alpha \dot{x}^\beta}}+qA_i.
\end{eqnarray}
From Eqs. \eqref{momentum} and~\eqref{momentumli2}, one can obtain
\begin{eqnarray}
\label{jacobimmei}
p_i\dot{x}^i&=&m\sqrt{\frac{\gamma_{ij}\dot{x}^i\dot{x}^j}{g_{\alpha\beta}\dot{x}^\alpha \dot{x}^\beta}}\sqrt{\gamma_{ij}\dot{x}^i\dot{x}^j}\nn\\
&&-\left(E+qA_t\right)\frac{g_{ti}}{g_{tt}}\dot{x}^i+qA_i\dot{x}^i,
\end{eqnarray}
where
\begin{eqnarray}
\gamma_{ij}=-g_{ij}+\frac{g_{ti}g_{tj}}{g_{tt}}.
\label{eq:gammaijdef}
\end{eqnarray}
In addition, Eq. \eqref{momentum} leads to the identity
\begin{eqnarray}
\label{momentum1}
m^2{g}_{tt}\left(1+\frac{\gamma_{ij}\dot{x}^i\dot{x}^j}{g_{\alpha\beta}\dot{x}^\alpha \dot{x}^\beta}\right)=\left(E+qA_t\right)^2.
\end{eqnarray}
Combining Eqs. \eqref{jacobimmei} to \eqref{momentum1}, the Jacobi metric $d\sigma=p_idx^i$ can be written as \cite{Chanda2019b}
\begin{eqnarray}
\label{fensleranders}
F(x,dx)&=&d\sigma(x,dx)\nn\\
&=&p_idx^i\nn\\
&=&\sqrt{\alpha_{ij}dx^idx^j}+\beta_idx^i,
\end{eqnarray}
where
\begin{eqnarray}
\label{fensleranders1}
&&\alpha_{ij}=\frac{\left(E+qA_t\right)^2-m^2{g}_{tt}}{{g}_{tt}}\gamma_{ij},\\
\label{fensleranders2}
&&\beta_i=qA_i-\left(E+qA_t\right)\frac{g_{ti}}{g_{tt}}.
\end{eqnarray}
Eqs. \eqref{fensleranders} to \eqref{fensleranders2} form a Randers type Finsler metric, where $\alpha_{ij}$ is a Riemanian metric and $\beta_i$ is a one-form, satisfying the positivity and convexity~\cite{Chern2002}
\begin{eqnarray}
\label{eq:fcond}
|\beta|=\sqrt{\alpha^{ij}\beta_i \beta_j}<1.
\end{eqnarray}
The Jacobi metric of charged particles given by Eqs. \eqref{fensleranders} to \eqref{fensleranders2} can be reduced in the neutral particle case by setting $q=0$, to~\cite{Chanda2019},
\begin{eqnarray}
\label{jacobineutra}
ds_J&=&\sqrt{\frac{E^2-m^{2} g_{t t}}{g_{t t}}\left(-g_{ij}+\frac{g_{ti}g_{tj}}{g_{tt}}\right)dx^idx^j}\nn\\
&&-E\frac{g_{ti}}{g_{tt}}dx^i.
\end{eqnarray}
Let $q=m=0$ and $E=1$, it further reduces to the optical metric,
\begin{eqnarray}
\label{opticallzh}
dt=\sqrt{\left(-\frac{g_{ij}}{g_{tt}}+\frac{g_{ti}g_{tj}}{g_{tt}^2}\right)dx^idx^j}-\frac{g_{ti}}{g_{tt}}dx^i.
\end{eqnarray}
In addition, when $g_{ti}=0$, Eqs. \eqref{fensleranders} to \eqref{fensleranders2}, Eq. \eqref{jacobineutra} and Eq. \eqref{opticallzh} correspond to the Jacobi metrics of charged particles, of neutral particles,  and the optical metric in a static spacetime, respectively.

Before proceeding further to study the deflection of charged particle, in the following two subsections we will respectively give the other two equivalent forms of the Randers form of the Jacobi metric Eqs. \eqref{fensleranders} to \eqref{fensleranders2}, in order to better understand the different methods used in Sec.~\ref{anglechen3}.

\subsection{Zermelo navigation problem}
In 1931, Zermelo considered a time-optimal control problem, which is to solve the shortest time path of particles moving in Euclidean space and affected by a vector field~\cite{Zermelo1931}. For a Riemannian metric $h_{ij}$ and a time independent vector field $W^i$ (wind), Shen showed that a natural solution of Zermelo navigation problem is the Randers metric~\cite{shenzhongmin2003}. In general, one can obtain the Randers data $(\alpha_{ij},\beta_i)$ from Zermelo data $(h_{ij},W^i)$ by the following transformation~\cite{Gibbons2009}
\begin{eqnarray}
&&\alpha_{ij}=\frac{\lambda h_{ij}+W_iW_j}{\lambda^2}, \\
&& \beta_i=-\frac{W_i}{\lambda}.
\end{eqnarray}
where
\begin{eqnarray}
&&\lambda=1-h_{ij}W^i W^j,~~~~W_i=h_{ij}W^j.
\end{eqnarray}
Conversely, for a Randers data $(\alpha_{ij},\beta_i)$, there is its corresponding Zermelo data~\cite{Gibbons2009}
\begin{eqnarray}
\label{Zermelo1}
&&h_{ij}=\lambda \left(\alpha_{ij}-\beta_i \beta_j\right), \\
\label{Zermelo2}
&& W^i=-\frac{\beta^i}{\lambda}.
\end{eqnarray}
where
\begin{eqnarray}
\label{Zermelo3}
&&\lambda=1-\alpha^{ij}\beta_i \beta_j,~~~~\beta^i=\alpha^{ij}\beta_j.
\end{eqnarray}
In short, Randers data $(\alpha_{ij},\beta_i)$ and Zermelo data $(h_{ij},W^i)$ are equivalent,

\subsection{$(n+1)$DSS picture}
Gibbons et al. proposed another equivalent viewpoint, namely the geodesic flow in an $n$-dimensional Randers space can be regarded as the null geodesic flow in an $(n+1)$DSS. Given an $n$-dimensional Randers space $(\alpha_{ij},\beta_i)$, this $(n+1)$DSS can be constructed as ~\cite{Gibbons2009}
\begin{eqnarray}
\label{lizongxiong}
d\hat{s}^2=\hat{g}_{ij}dx^idx^j=V^2\left[\left(dt-\beta_idx^i\right)^2-\alpha_{ij}dx^i dx^j\right],~~~
\end{eqnarray}
where $V$ is a conformation factor. Since null geodesics are conformally invariant, the choice of $V$ is very arbitrary. Considering Jacobi-Randers metric given by Eqs. \eqref{fensleranders} to \eqref{fensleranders2}, it is more convenient to choose
\begin{eqnarray}
&&V^{2}=\frac{{g}_{tt}}{\left(E+qA_t\right)^2-m^2{g}_{tt}}.
\end{eqnarray}

One can verify that for the optical metric  of spcetime~\eqref{lizongxiong}, we have
\begin{eqnarray}
dt=\sqrt{\alpha_{ij}dx^idx^j}+\beta_idx^i,
\end{eqnarray}
which is the same as the Jacobi metric $F(x,dx)$ given in Eq. ~\eqref{fensleranders}. Since the time $t$ in above equation is not the physics time, following the idea of Ref.~\cite{Crisnejodengren}, one can define a new Jacobi metric based on $F(x,dx)$, as follows
\begin{eqnarray}
\label{zhudongmei}
\tilde{F}(x,dx)&\equiv &\frac{F(x,dx)}{E}\nn\\
&=&\frac{1}{E}\left(\sqrt{\alpha_{ij}dx^idx^j}+\beta_idx^i\right)\nn\\
&=&\sqrt{\tilde{\alpha}_{ij}dx^idx^j}+\tilde{\beta}_idx^i,
\end{eqnarray}
where
\begin{eqnarray}
\label{newaobama1}
&&\tilde{\alpha}_{ij}=\frac{\alpha_{ij}}{E^2}=\frac{\left(1+\frac{qA_t}{E}\right)^2-\left(\frac{m}{E}\right)^2{g}_{tt}}{{g}_{tt}}\gamma_{ij},\\
\label{newaobama2}
&&\tilde{\beta}_i=\frac{\beta_i}{E}=\frac{qA_i}{E}-\left(1+\frac{qA_t}{E}\right)\frac{g_{ti}}{g_{tt}}.
\end{eqnarray}
In terms of the new Randers data $(\tilde{\alpha}_{ij},\tilde{\beta}_i)$, the $(n+1)$DSS metric \eqref{lizongxiong} can be rewritten as
\begin{eqnarray}
\label{wangqinzhuo}
d\tilde{s}^2=\tilde{g}_{ij}dx^idx^j=\tilde{V}^2\left[\left(dt-\tilde\beta_idx^i\right)^2-\tilde\alpha_{ij}dx^i dx^j\right],~~~
\end{eqnarray}
where
\begin{eqnarray}
\tilde{V}^2=\frac{{g}_{tt}}{\left(1+\frac{qA_t}{E}\right)^2-\left(\frac{m}{E}\right)^2{g}_{tt}}.
\end{eqnarray}
In this article, we will take the viewpoints of Ref.~\cite{Crisnejodengren} to directly correspond the geodesic motion of charged particles in the $n$-dimensional Jacobi Randers space to the null geodesic motion in the $(n+1)$DSS described by Eq. \eqref{wangqinzhuo}.

\subsection{Gauss-Bonnet theorem and deflection angle formulas}
We assume that the trajectory of the particle lies in a two-dimensional space called lensing geometry. In this subsection, the GB theorem will be applied to the lensing geometry, and the formulas for calculating the deflection angle using curvature are obtained.

Let $D$ be a subset of a compact, oriented surface with Gaussian curvature $K$ and Euler characteristic number $\chi(D)$, and its boundary $\partial{D}$ be a piecewise smooth curve with geodesic curvature $k$. The GB theorem states~\cite{Carmo1976}
\begin{equation}
\label{GBT}
\iint_{D}{K}dS+\oint_{\partial{D}}k~dl+\sum_{i=1}{\varphi_i}=2\pi\chi({D}),
\end{equation}
where $dS$ is the area element, $dl$ is the line element of boundary, and $\varphi_i$ is the jump angle in the $i$-th vertex of $\partial{D}$ in the positive sense, respectively.

\begin{figure}[htp!]
\label{Figure}
\centering
\includegraphics[width=8.0cm]{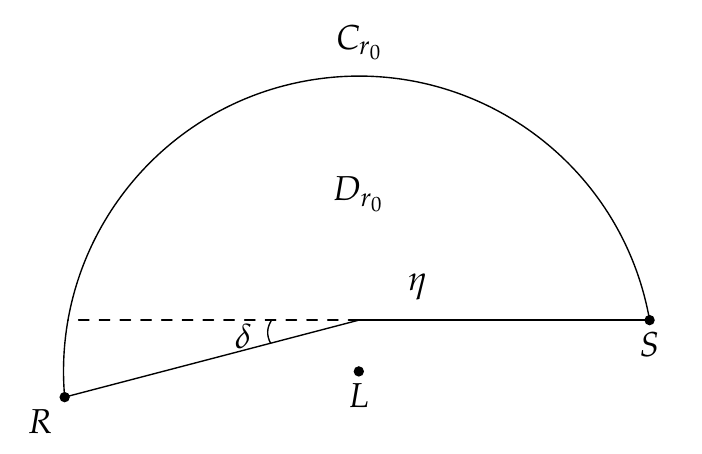}
\caption{A region $D_{r_0}\subset(\mathcal{M},g^L_{ij})$ with boundary $\partial D_{r_0}=\eta \cup C_{r_0}$. $S$, $R$ and $L$ denote the source, the receiver, and the lens, respectively. $\delta$ is the deflection angle.\label{fig:1}}
\end{figure}

This can be applied to the lensing geometry $(\mathcal{M},g^L_{ij})$ with coordinates ($r,\phi$), which contains $D_{r_0}$ (see Fig. \ref{fig:1}), a non singular and asymptotically Euclidean region. The boundary of the region $\partial D_{r_0}=\eta \cup C_{r_0}$, where $\eta$ is the particle ray from the source $S$ to the receiver $R$, and $C_{r_0}$ is a curve defined by $r=r_0=\mathrm{constant}$. We can see that $\chi(D_{r_0})=0$ because $D_{r_0}$ is a non singular region.

The Gaussian curvature of $(\mathcal{M},g^L_{ij})$ can be calculated by following equation~\cite{Werner2012}
\begin{eqnarray}
\label{Gausscurvature}
K=\frac{1}{\sqrt{\det g^L}}\left[\frac{\partial \left(\frac{\sqrt{\det g^L}}{g^L_{rr}}{{\Gamma}^\phi_{rr}}\right)}{\partial{\phi}}-\frac{\partial\left(\frac{\sqrt{\det g^L}}{g^L_{rr}}{{\Gamma}^\phi_{r\phi}}\right)}{\partial{r}}\right],~~~
\end{eqnarray}
where $\det g^L$ denotes the determinant of metric $g^L_{ij}$.
For the geodesic curvature part, when $r_0\rightarrow \infty$, we have $k(C_{r_0})dl\rightarrow d\phi$, and therefore $\int_{C_{\infty}}k(C_{\infty}) dl=\int_0^{\pi+\delta}d\phi$, with $\delta$ the asymptotic deflection angle. And for the jump angles in $S$ and $R$, denoted as $\varphi_S$ and $\varphi_R$ respectively, we see that $\varphi_R+\varphi_S\rightarrow\pi$ as $r_0\rightarrow\infty$. Putting them together according to Eq. \eqref{GBT}, we have
\begin{eqnarray}
&&\iint_{D_{r_0}}K dS-\int_S^R \kappa (\eta)dl+\int_{\phi_S}^{\phi_R}d\phi+\varphi_R+\varphi_S\nn\\
&\stackrel{r_0\rightarrow\infty}{=}&\iint_{D_{\infty}}K  dS-\int_S^R \kappa (\eta)dl+\int_0^{\pi+\delta}d\phi+\pi\nn\\
&=&2\pi.
\end{eqnarray}
From this, we can solve the deflection angle as
\begin{eqnarray}
\label{OIAojacobi}
\delta&=&-\iint_{D_\infty}{K } dS+\int_S^R \kappa (\eta)dl.
\end{eqnarray}
In particular, when the particle ray $\eta$ is a geodesic,  $\kappa(\eta)=0$, the deflection angle simplifies to
\begin{eqnarray}
\label{Werner}
\delta&=&-\iint_{D_\infty}{K } dS.
\end{eqnarray}

\section{Kerr-Newman Jacobi geometry for charged particle in three forms}\label{KNjacobi}
In this section, the Jacobi metric in three equivalent descriptions introduced in Sec.~\ref{jacobimetric} will be specifically applied to KN spacetime, to prepare for the computation of deflection angle using different methods in the next section.

The line element of KN spacetime in Boyer-Lindquist coordinates reads~\cite{newman1,newman2}
\begin{eqnarray}
\label{knmetric}
ds^2&=&\left(1-\frac{2Mr-Q^2}{\Sigma}\right)dt^2-\frac{\Sigma}{\Delta}dr^2-\Sigma d\theta^2\nn\\
&&-\sin^2\theta \left(r^2+a^2+\frac{(2Mr-Q^2)}{\Sigma}a^2\sin^2\theta\right)d\phi^2\nn\\
&&+2a\sin^2\theta \frac{2M r-Q^2}{\Sigma}d\phi dt,
\end{eqnarray}
where
\begin{eqnarray}
&&\Sigma=r^2+a^2\cos^2\theta,~~\Delta=r^2-2Mr+a^2+Q^2,\nn
\end{eqnarray}
and, $M$, $Q$ and $a$ are the mass, charge and angular momentum per unit mass of the black hole, respectively. Its gauge field is
\begin{eqnarray}
\label{gaugefield}
&&A_\mu dx^\mu=\frac{Qr}{\Sigma}\left(dt-a\sin^2\theta d\phi\right).
\end{eqnarray}

\subsection{KN Jacobi-Randers metric}

Substituting KN metric~\eqref{knmetric} and gauge field~\eqref{gaugefield} into Eqs. \eqref{eq:gammaijdef} and \eqref{fensleranders} to \eqref{fensleranders2}, the KN Jacobi metric for a charged particle in Randers form can be written as
\begin{eqnarray}
\label{KNranders}
&&F(x,dx)=d\sigma=\sqrt{\alpha_{ij}d x^{i} d x^{j}}+\beta_i d x^i,\nn\\
&&\alpha_{ij}=V^{-2}\gamma_{ij},~~~~\beta_i d x^i=\beta_\phi d \phi,
\end{eqnarray}
where $V^{-2},~\gamma_{ij},~\beta_\phi$ are given by
\begin{eqnarray}
\label{KNranders2}
V^{-2}&=&\frac{\left(E-\frac{qQr}{\Sigma}\right)^2-m^2\frac{\Delta-a^2\sin^2\theta}{\Sigma}}{\frac{\Delta-a^2\sin^2\theta}{\Sigma}},\\
\label{KNranders3}
\gamma_{i j} d x^{i} d x^{j}&=&\frac{\Sigma}{\Delta} d r^{2}+\Sigma d \theta^{2}+\frac{\sin^2\theta}{\Sigma}\bigg[\left(a^{2}+r^{2}\right)^{2}\nn\\
&&-a^{2} \Delta \sin^2\theta+\frac{a^{2}\left(2 M r-Q^{2}\right)^{2} \sin^2 \theta}{\left(\Delta-a^{2} \sin^2\theta\right)}\bigg] d \phi^{2},\nn\\
\\
\label{KNranders4}
\beta_{\phi}&=&\frac{a \sin^2 \theta}{\Sigma\left(a^{2} \sin^2 \theta-\Delta\right)}\bigg[-q Q r^{3}+E\left(r^{2}-\Delta\right) \Sigma\nn\\
&&+a^{2}\left(E\Sigma-q Q r\right)+a^{2} qQr \sin^2 \theta\bigg].
\end{eqnarray}
The Finsler condition \eqref{eq:fcond} becomes
\begin{eqnarray}
|\beta|^2&=&\frac{a^{2}\left[q Q r+E\left(Q^{2}-2 M r\right)\right]^{2} \sin^2\theta}{\left[\left(q Q r-E\Sigma\right)^{2}-m^2\Sigma\left(\Delta-a^{2} \sin^2\theta\right)\right] \Delta}<1.\nn
\end{eqnarray}

\subsection{KN Jacobi metric in Zermelo form}

From Randers data $(\alpha_{ij},\beta_i)$ given by Eqs. \eqref{KNranders} to \eqref{KNranders4}, one can write the KN Jacobi metric in the Zermelo data form via transformation~\eqref{Zermelo1} to \eqref{Zermelo3}, as follows
\begin{eqnarray}
\label{KNZermelo}
&&h_{ij}dx^idx^j=\left(1-|\beta|^2\right) \left(V^{-2}\gamma_{ij}dx^idx^j-\beta_\phi^2d\phi^2\right),\nn\\
&& W^i\frac{\partial}{\partial x^i}=-\frac{a\left[qQr+E\left(Q^2-2Mr\right)\right]}{\left(1-|\beta|^2\right)V^{-2}\Delta\Sigma}\frac{\partial}{\partial \phi}.
\end{eqnarray}
Let $q=m=0$ and $E=1$, the optical Zermelo metric reduces to
\begin{eqnarray}
&&h_{ij}dx^idx^j=\frac{H(r,\theta)}{\Delta}\left(\frac{d r^{2}}{\Delta}+d \theta^{2}+\frac{H(r,\theta)\sin^2\theta}{\Sigma^{2}} d \phi^{2}\right),\nn\\
&& W^i\frac{\partial}{\partial x^i}=\frac{a\left(2  M r-Q^2\right)}{H(r,\theta)}\frac{\partial}{\partial \phi},
\end{eqnarray}
where
\begin{eqnarray}
&&H(r,\theta)\equiv \left(a^{2}+r^{2}\right)^{2}-a^{2} \Delta \sin^2\theta.\nn
\end{eqnarray}
For Kerr black hole, $Q=0$ and the optical Zermelo metric becomes
\begin{eqnarray}
&&h_{ij}dx^idx^j=\frac{H(r,\theta)}{\Delta_K}\left(\frac{d r^{2}}{\Delta_K}+d \theta^{2}+\frac{H(r,\theta)\sin^2\theta}{\Sigma^{2}} d \phi^{2}\right),\nn\\
&& W^i\frac{\partial}{\partial x^i}=\frac{2a M r}{H(r,\theta)}\frac{\partial}{\partial \phi},
\end{eqnarray}
where $\Delta_K=r^2-2Mr+a^2$. This equation agrees with Eq. (97) of Ref.~\cite{Gibbons2009}.

\subsection{$(n+1)$DSS picture of the KN Jacobi geometry}

Substituting Randers data $(\alpha_{ij},\beta_i)$ in Eqs. \eqref{KNranders} to \eqref{KNranders4} into Eqs. \eqref{newaobama1} and~\eqref{newaobama2}, the new Randers data $(\tilde\alpha_{ij},\tilde\beta_i)$ can be obtained. Then one can use the 3-dimensional Randers data $(\tilde\alpha_{ij},\tilde\beta_i)$ to write the $(3+1)$DSS form of Jacobi metric based on Eq. \eqref{wangqinzhuo}, i.e.,
\begin{eqnarray}
\label{wangqinzhuo1}
d\tilde{s}^2&=&\tilde{V}^2\left[\left(dt-\tilde\beta_idx^i\right)^2-\tilde{\alpha}_{ij}dx^i dx^j\right].
\end{eqnarray}
Since the purpose of this article is to calculate the second-order deflection angle, one can expanded the components of metric~\eqref{wangqinzhuo1} as power series of $1/r$. For simplicity, this article only considers motion in the equatorial plane $(\theta=\frac{\pi}{2})$, and then the result of the components of this metric reads
\begin{eqnarray}
\label{spacetime1}
\tilde{g}_{tt}&=&\frac{1}{v^{2}}-\frac{2 M}{r v^{4}}+\frac{2 q Q \sqrt{1-v^{2}}}{m r v^{4}}+\frac{4 M^{2}\left(1-v^{2}\right)}{r^{2} v^{6}}\nn\\
&&-\frac{4 M q Q \sqrt{1-v^{2}}\left(2-v^{2}\right)}{m r^{2} v^{6}}+\frac{Q^{2}}{r^{2} v^{4}}\nn\\
&&+\frac{q^{2} Q^{2}\left(4-5 v^{2}+v^{4}\right)}{m^{2} r^{2} v^{6}}+ \mathcal{O}\left(\frac{[M]^3}{r^3}\right), \\
\label{spacetime2}
\tilde{g}_{rr}&=&-\left(1+\frac{2 M}{r}+\frac{4 M^{2}}{r^{2}}-\frac{Q^{2}+a^2}{r^{2}}\right)\nn\\
&&+ \mathcal{O}\left(\frac{[M]^3}{r^3}\right), \\
\label{spacetime3}
\tilde{g}_{\phi\phi}&=&-\left(r^{2}+a^2\right)+ \mathcal{O}\left(\frac{[M]^3}{r}\right),  \\
\label{spacetime4}
\tilde{g}_{t\phi}&=&\frac{2 M a}{r v^{2}}-\frac{a q Q \sqrt{1-v^{2}}}{m r v^{2}}+ \mathcal{O}\left(\frac{[M]^3}{r^2}\right),
\end{eqnarray}
in which we have used
\begin{eqnarray}
E=\frac{m}{\sqrt{1-v^2}},~~~~
\end{eqnarray}
with $v$ being the asymptotic velocity of the charged particle. Here and henceforth, in the higher order corrections we use $[M]^n$ to collectively denote products of $\{M,Q,a,q,m^{-1}\}$ with dimension $M^n$. For example, $[M]^3$ might include terms proportional to $M^3,~M^2Q,~MQ^2,~\cdots,~aqQ^2/m,\cdots$ etc.

\section{Deflection angle of charged particle by a Kerr-Newman black hole}\label{anglechen3}
In this section we will calculate the second-order deflection angle of charged particle moving in the equatorial plane of KN spacetime, using the Randers data and the $(n+1)DSS$ picture presented in the previous two sections. For the KN Jacobi-Randers geometry, we will use two methods utilizing the Gauss-Bonnet theorem. The first one is Werner's ORMM and the other one is the GJMM. Using the $(n+1)$DSS picture, we can calculate the deflection angle by null geodesics.

Because particle orbits are required regardless of the method, we will first consider the geodesic method which solves the deflection angle iteratively. It should be noted that applying the GB theorem to calculate the second-order deflection angle requires only first order  orbit information.

\subsection{Geodesic method using iteration}
The spiritual essence of this subsection is to correspond the non-geodesic motion of charged particles in KN spacetime with gauge filed or the geodesic motion of particle in $3$-dimensional Randers space to geodesic motion of light in $(3+1)$DSS. We can use the geodesic equation of photons to calculate the deflection angle. Considering the equatorial plane $\theta=\pi/2$, the Lagrangian of a photon in $(3+1)$DSS described by metric \eqref{wangqinzhuo1} is
 \begin{eqnarray}
\label{r1}
 2L=\tilde{g}_{\mu \nu} \dot{x}^{\mu} \dot{x}^{\nu}=\tilde{g}_{tt}\dot{t}^2+2\tilde{g}_{t\phi}\dot{t } \dot{\phi}+\tilde{g}_{rr}\dot{r}^2+\tilde{g}_{\phi\phi}\dot{\phi}^2~.~~~~
\end{eqnarray}
Then one can obtain its conserved energy $E$ and conserved angular momentum $J$,
\begin{eqnarray}
\label{liangyumo1}
&& \left(\tilde{g}_{tt}\dot{t}+\tilde{g}_{t\phi}\dot{\phi}\right)=E~,\\
\label{liangyumo2}
&&-\left(\tilde{g}_{t\phi}\dot{t}+\tilde{g}_{\phi\phi}\dot{\phi}\right)=J~.
\end{eqnarray}
Combining Eqs. \eqref{r1}-\eqref{liangyumo2} and considering null condition $L=0$, one can obtain the following orbit equation
\begin{eqnarray}
\label{hawking}
\left(\frac{d r}{d \phi}\right)^{2}=\frac{\left(b^2 v^{2} \tilde{g}_{tt}+2 b v \tilde{g}_{t\phi}+\tilde{g}_{\phi\phi}\right)\left(\tilde{g}_{tt} \tilde{g}_{\phi\phi}-\tilde{g}_{t\phi}^{2}\right)}{\left(b v \tilde{g}_{tt}+\tilde{g}_{t\phi}\right)^{2} \tilde{g}_{rr}}.
\end{eqnarray}
Note that we have use $bv=J/E$, with $b$ being the impact parameter. Using the metric component of 3-spacetime given by Eqs. \eqref{spacetime1}-\eqref{spacetime4}, this orbit equation can be solved using the iteration method, as show in Appx. \ref{liuhaotian}. In this case, the calculation of the deflection angle is very straightforward and intuitive. The result of the deflection angle is (see Eq. \eqref{eq:angleapp})
\begin{eqnarray}
\label{anglespicture}
 \delta&=&2\left( 1+\frac{1}{v^2} -\frac{\hat{q}\hat{Q}\sqrt{1-v^2}}{v^2}\right)\frac{M}{b}\nn\\
 &&+\bigg\{3\pi\left( \frac{1}{4}+\frac{1}{v^2} \right)-\pi\left( \frac{1}{4}+\frac{1}{2v^2}\right) \hat{Q}^2-\frac{4 \hat{a}}{v}\nn\\
 &&-\frac{3\pi}{v^2}\hat{q}\hat{Q}\sqrt{1-v^2}\nn+\frac{2\hat{a}}{v}\hat{q}\hat{Q}\sqrt{1-v^2}\\
 &&+\frac{\pi}{2v^2} \hat{q}^2\hat{Q}^2\left(1-v^2\right) \bigg\}\frac{M^2}{b^2}\nn\\
 &&+ \mathcal{O}\left(\frac{[M]^3}{b^3}\right),
\end{eqnarray}
where the charge-to-mass ratio $\hat{q}=q/m$ and $\hat{Q}=Q/M$ and $\hat{a}=a/M$.

\subsection{ORMM using Randers data}
With Randers data $(\alpha_{ij},\beta_i)$, this and the next subsections will use the GB theorem to study the deflection of charged particle. We consider Werner' ORMM first.

The Hessian of a Finsler metric $F(x,y)$ of a smooth manifold $M$ reads~\cite{Chern2002}
\begin{eqnarray}
\label{Hessian}
 g_{ij}(x,y)&=&\frac{1}{2}\frac{\partial^2F^2(x,y)}{\partial y^i \partial y^j}~,
\end{eqnarray}
where $(x,y)\in T_M$ with $T_M$ being the tangent bundle of $M$.
In Ref.~\cite{Werner2012}, Werner applied Naz{\i}m's method to construct an  osculating Riemannian manifold $(M, \bar{g})$ of Finsler manifold $(M, F)$. Following~\cite{Werner2012}, one can choose a smooth nonzero vector field $Y$ tangent to the geodesic $\eta_F$, i.e. $Y(\eta_F)=y$, and thus the osculating Riemannian metric can be obtained from the Hessian and the geodesic
\begin{eqnarray}
\label{lingling}
 \bar{g}_{ij}(x)&=&g_{ij}\left(x,Y(x)\right)~.
\end{eqnarray}
In this construction, the geodesic $\eta_F$ of $(M, F)$ is also a geodesic $\eta_{\bar{g}}$ of $(M, \bar{g})$ . On the equatorial plane $(\theta=\pi/2)$, the Finsler metric of Randers type given by Eqs. \eqref{KNranders} to \eqref{KNranders4} leads to
\begin{eqnarray}
\label{Randers-FinslerY}
F\left(r,\phi,Y^r,Y^\phi\right)=\sqrt{\alpha_{ij}(r,\phi)Y^i Y^j}+\beta_\phi(r,\phi) Y^\phi ~,
\end{eqnarray}
where
\begin{eqnarray}
\label{Randers-FinslerY1}
&&\alpha_{ij}Y^iY^j=V^{-2}\left[\frac{r^{2}}{\Delta} (Y^r)^{2}+\frac{r^{2} \Delta}{\Delta-a^{2}} (Y^\phi)^{2}\right],~~~~~~~~\\
\label{Randers-FinslerY2}
&&\beta_{\phi}=\frac{a\left[\frac{m}{\sqrt{1-v^{2}}}\left(Q^{2}-2 M r\right)+q Q r\right]}{\Delta-a^{2}},\\
\label{Randers-FinslerY3}
&&V^{-2}=\frac{\left(q Q-\frac{m r}{\sqrt{1-v^{2}}}\right)^{2}}{\Delta-a^{2}}-m^{2}.
\end{eqnarray}
Considering the zeroth-order particle ray $r=b/\sin\phi$ (see Eq. \eqref{eq:rnres}), one can choose the following vector fields (see Werner~\cite{Werner2012} for a detailed discussion)
\begin{eqnarray}
\label{niubidun1}
&&Y^r=\frac{dr}{d\sigma}=-\frac{\sqrt{1-v^2}\cos\phi}{m  v},\\
\label{niubidun2}
&&Y^\phi=\frac{d\phi}{d\sigma}=\frac{\sqrt{1-v^2}\sin^2\phi}{ b v m}~.
\end{eqnarray}
Now substituting Eqs. \eqref{Randers-FinslerY}-\eqref{Randers-FinslerY3} into Eq. \eqref{Hessian}, the Hessian can be obtained in terms of $Y^r$ and $Y^\phi$. And substituting this together with Eqs. \eqref{niubidun1} and~\eqref{niubidun2} into Eq. \eqref{lingling}, the metric of the osculating Riemannian metric can be found. Because of its excessive length, we only list its components in Appx. \ref{omga}.

Since the particle ray is a geodesic in $(M,\bar{g}_{ij})$ and $D_{r_0}\subset(M,\bar{g}_{ij})$, the deflection angle can be calculated by Eq. \eqref{Werner}, written in more detail as
\begin{eqnarray}
\label{lizonghai2}
\delta=-\iint_{D_\infty} \bar{K}dS=-\int_0^{\pi}\int_{r(\phi)}^{\infty}\bar{K}\sqrt{\det{\bar{g}}}drd\phi,
\end{eqnarray}
where $\bar{K}$ is the Gaussian curvature of the osculating Riemannian metric and can be computed by substituting Eqs. \eqref{wuhandaxue1} to \eqref{wuhandaxue3} into Eq. \eqref{Gausscurvature}. To order $\mathcal{O}(1/r^4)$,  the Gaussian curvature is found to be
\begin{align}
\label{gaussgauss}
\bar{K}=&-\frac{3}{2v} \left(1-\frac{1}{v^2}\right)\left( 2-\hat{q}\hat{Q} \sqrt{1-v^2}\right) f(r, \phi) \frac{a }{b^2}\frac{ M}{m^2 r^2}\nn\\
&+\left(1-\frac{1}{v^2}\right)\left[1+\frac{1}{v^2}- \frac{\hat{q} \hat{Q}\sqrt{1-v^2}}{v^2}\right]\frac{M}{m^2 r^3}\nn\\
&+\left(1-\frac{1}{v^2}\right)\bigg[3\left(1-\frac{2}{v^2}\right)-3\left(1-\frac{4}{v^2}\right)\frac{\hat{q}\hat{Q}\sqrt{1-v^2}}{v^2} \nn\\
&- \left(1+\frac{2}{v^2}\right)\hat{Q}^2 +2 \left(1-\frac{3}{v^2}\right)\frac{\hat{q}^2  \hat{Q}^2 (1-v^2)}{v^2} \bigg] \frac{M^2}{m^2 r^4}\nn\\
&+ \mathcal{O}\left(\frac{[M]}{r^5}\right),
\end{align}
where \cite{Werner2012}
\begin{eqnarray}
&&f(r, \phi)=\frac{\sin^3\phi}{\left(\cos^2\phi+\frac{r^{2}}{b^{2}} \sin^4\phi\right)^{\frac{7}{2}}} \bigg[2 \cos^6 \phi\left(\frac{5r}{b} \sin\phi-2\right)~~~~~\nn\\
&&~~~~+\cos^4 \phi \sin^2 \phi\left(-2+9 \frac{r}{b} \sin \phi-10 \frac{r^{3}}{b^{3}} \sin^3 \phi\right)\nn\\
&&~~~~+4 \frac{r}{b} \cos^2\phi\sin^5 \phi\left(1+2 \frac{r}{b} \sin\phi-\frac{r^{2}}{b^{2}}\sin^2\phi\right)\nn\\
&&~~~~+\frac{r^{2}}{b^{2}}\left(-\frac{r}{b} \sin^9\phi+2 \frac{r^{3}}{b^{3}} \sin^{11} \phi+\sin^4(2 \phi)\right)\bigg].~~~~\nn
\end{eqnarray}
Using this and the first order  particle orbit in~Eq. \eqref{orbit}, the deflection angle can be obtained by Eq. \eqref{lizonghai2} and the result reads
\begin{eqnarray}
\label{anglerpicture}
 \delta&=&2\left( 1+\frac{1}{v^2} -\frac{\hat{q}\hat{Q}\sqrt{1-v^2}}{v^2}\right)\frac{M}{b}\nn\\
 &&+\bigg\{3\pi\left( \frac{1}{4}+\frac{1}{v^2} \right)-\pi\left( \frac{1}{4}+\frac{1}{2v^2}\right) \hat{Q}^2-\frac{4 \hat{a}}{v}\nn\\
 &&-\frac{3\pi}{v^2}\hat{q}\hat{Q}\sqrt{1-v^2}\nn+\frac{2\hat{a}}{v}\hat{q}\hat{Q}\sqrt{1-v^2}\\
 &&+\frac{\pi}{2v^2} \hat{q}^2\hat{Q}^2\left(1-v^2\right) \bigg\}\frac{M^2}{b^2} + \mathcal{O}\left(\frac{[M]^3}{b^3}\right),
\end{eqnarray}
One can find that Eq. \eqref{anglerpicture} is in perfect agreement with Eq. \eqref{anglespicture}. Seting $q=0$, the result reduces to the deflection angle for neutral particles in KN spacetime~\cite{Helin2016}. And setting $a=0$, it leads to the deflection angle of charged particles in Reissner-Nordstr\"{o}m spacetime~\cite{lidujia,Xu:2021rld}.

\subsection{GJMM using Randers data}
In this subsection, we still use Randers data $(\alpha_{ij},\beta_i)$. However, the particle now is supposed moving in Riemanian space described by $\alpha_{ij}$
 \begin{eqnarray}
\label{lineelement}
dl^2=\alpha_{ij}dx^idx^j,
\end{eqnarray}

In the spirit of GJMM, the motion of the particles no longer follows the geodesic. Use $(M,\alpha_{ij})\supset D_{\infty}$ as the lensing geometry, the deflection angle can be calculated by Eq. \eqref{OIAojacobi}. That is,
\begin{eqnarray}
\label{kunlunjue}
\delta&=&\iint_{D_\infty} K dS+\int_{S}^{R} \kappa(\eta) dl\nn\\
&=&-\int_{0}^{\pi} \int_{r(\phi)}^{\infty}K \sqrt{\det \alpha} dr d\phi+\int_{0}^{\pi} \kappa(\eta) \frac{dl}{d\phi} d\phi\nn\\
&\equiv& \delta_{gau}+\delta_{geo},\label{gausspart}
\end{eqnarray}
where in the last step we have split $\delta$ into the Gaussian curvature part $\delta_{gau}$, i.e. the first term,  and the geodesic curvature part $\delta_{geo}$, the second term.

The Gaussian curvature of $\alpha_{ij}$ can be obtained by substituting Eqs. \eqref{KNranders} to \eqref{KNranders3} into Eq. \eqref{Gausscurvature} and reduce it in equatorial plane $(\theta=\pi/2)$. The result is
\begin{eqnarray}
\label{gaussZermelo}
K&=&\left(1-\frac{1}{v^2}\right)\left[1+\frac{1}{v^2}- \frac{\hat{q} \hat{Q}\sqrt{1-v^2}}{v^2}\right]\frac{M}{m^2 r^3}\nn\\
&&+\left(1-\frac{1}{v^2}\right)\bigg[3\left(1-\frac{2}{v^2}\right)\nn\\
&&-3\left(1-\frac{4}{v^2}\right)\frac{\hat{q}\hat{Q}\sqrt{1-v^2}}{v^2} - \left(1+\frac{2}{v^2}\right)\hat{Q}^2\nn\\
&&+2 \left(1-\frac{3}{v^2}\right)\frac{\hat{q}^2  \hat{Q}^2 (1-v^2)}{v^2} \bigg] \frac{M^2}{m^2 r^4}\nn\\
&&+ \mathcal{O}\left(\frac{[M]}{r^5}\right),
\end{eqnarray}
Substituting this and the first order particle ray in Eq. \eqref{orbit} into Eq. \eqref{gausspart}, the Gaussian curvature part reads
\begin{eqnarray}
 \delta_{gau}&=&2\left( 1+\frac{1}{v^2} -\frac{\hat{q}\hat{Q}\sqrt{1-v^2}}{v^2}\right)\frac{M}{b}\nn\\
 &&+\bigg\{3\pi\left( \frac{1}{4}+\frac{1}{v^2} \right)-\pi\left( \frac{1}{4}+\frac{1}{2v^2}\right) \hat{Q}^2\nn\\
 &&-\frac{3\pi}{v^2}\hat{q}\hat{Q}\sqrt{1-v^2}+\frac{\pi}{2v^2} \hat{q}^2\hat{Q}^2\left(1-v^2\right) \bigg\}\frac{M^2}{b^2} \nn\\
 &&+ \mathcal{O}\left(\frac{[M]^3}{b^3}\right), \label{eq:dgaussres}
\end{eqnarray}

On the other hand, the geodesic curvature of particle ray can be calculated by the following equation~\cite{OIA2017}
\begin{eqnarray}
&&\kappa(\eta)=-\frac{\beta_{\phi, r}}{\sqrt{(\det\alpha)\alpha^{\theta \theta}}}\Bigg|_{\theta=\pi/2}.
\end{eqnarray}
Using 3-space metric $\alpha_{ij}$ and one-form $\beta_i$ in Eqs. \eqref{KNranders} to \eqref{KNranders4} in this, one can obtain $\kappa(\eta)$ to the order $\mathcal{O}\left([M]/r^3\right)$ as
\begin{eqnarray}
\label{piaoliuping}
\kappa(\eta)=\bigg[-2+\hat{q} \hat{Q} \sqrt{1-v^2}\bigg]\frac{\hat{a} \sqrt{1-v^2}}{m v^2}\frac{M^2}{r^3}+ \mathcal{O}\left(\frac{[M]^2}{r^4}\right).
\end{eqnarray}
Using this, together with the line element Eq. \eqref{lineelement}, first order  particle ray in Eq. \eqref{orbit}, the geodesic curvature part of the geodesic angle can also be computed using Eq. \eqref{gausspart}. The result is found to be
\begin{eqnarray}
\delta_{geo}=2\left[-2+\hat{q}\hat{Q}\sqrt{1-v^2}\right]\frac{\hat{a}}{v}\frac{M^2}{b^2}+ \mathcal{O}\left(\frac{[M]^3}{b^3}\right). \label{eq:dgeores}
\end{eqnarray}

Finally, combining Eqs. \eqref{eq:dgaussres} and \eqref{eq:dgeores}, one can verify that the total deflection angle $\delta=\delta_{gau}+\delta_{geo}$ is consistent with the result Eq. \eqref{anglespicture} obtained by calculation of null geodesic in $(3+1)$-dimensional stationary spacetime, and the result Eq. \eqref{anglerpicture} obtained by Werner's ORMM in 3-dimensional Randers space. It is also interesting to note that the deflection caused by spin of the spacetime is and only is present in the geodesic curvature part $\delta_{geo}$ and $\delta_{geo}$ only contains terms involving the spacetime spin.

\subsection{Discussion of results}\label{result}
The second-order deflection angle of charged particle in the equatorial plane of KN spacetime obtained by the three methods are the same, which are quoted as
\begin{eqnarray}
\label{asymptoticangle}
\delta&=&2\left( 1+\frac{1}{v^2} -\frac{\hat{q}\hat{Q}\sqrt{1-v^2}}{v^2}\right)\frac{M}{b}\nn\\
 &&+\bigg\{3\pi\left( \frac{1}{4}+\frac{1}{v^2} \right)-\pi\left( \frac{1}{4}+\frac{1}{2v^2}\right) \hat{Q}^2-\frac{4 \hat{a}}{v}\nn\\
 &&-\frac{3\pi}{v^2}\hat{q}\hat{Q}\sqrt{1-v^2}+\frac{2\hat{a}}{v}\hat{q}\hat{Q}\sqrt{1-v^2}\nn\\
 &&+\frac{\pi}{2v^2} \hat{q}^2\hat{Q}^2\left(1-v^2\right) \bigg\}\frac{M^2}{b^2}\nn\\
 &&+ \mathcal{O}\left(\frac{[M]^3}{b^3}\right).
\end{eqnarray}
This article assumes that $b>0$ if the trajectory initially rotate counter-clockwisely around the center but the spacetime spin $a$ can be both positive or negative.

The result \eqref{asymptoticangle} is the deflection angle measured by receiver at spacial infinity for rays from source also at infinity. Recently, the finite distance  effects on deflection angle has attracted the interest of some authors~\cite{ISOA2016,IOA2017,Lijia2020,LiA2020,LizhouRas,lidujia}. In this paper, we mainly focus on the concepts and methodology, whereas puts the computationally more complicated finite distance  deflection in Appx. \ref{finidis} for readers' reference. When the distance between the source and the receiver from the KN lens tends to infinity, the finite distance  deflection angle \eqref{qiaofeng} can leads to the asymptotic deflection angle \eqref{asymptoticangle}.

In Ref.~\cite{Jusufi:cmp}, Jusufi used the GB theorem and a Riemannian optical metric to obtain the following result (see Eq. (32) of Ref.~\cite{Jusufi:cmp})
\begin{eqnarray}
\label{jusufiasymptoticangle}
\delta&=&2 \left( 1+\frac{1}{v^2}-\frac{\hat{q} \hat{Q}\sqrt{1-v^2}}{v^2} \right)\frac{M}{b}\nn\\
&&-\left[\pi\left(\frac14+\frac{1}{2v^2} \right) \hat{Q}^2-\frac{4\hat{a}}{v}\right]\frac{M^2}{b^2}+ \mathcal{O}\left(\frac{[M]^3}{b^3}\right).
\end{eqnarray}
Comparing Eq. \eqref{asymptoticangle} with \eqref{jusufiasymptoticangle}, one can find that the following new terms appear in our deflection angle
\begin{eqnarray}
&&\delta_{M^2}=3\pi\left( \frac14+\frac{1}{v^2} \right) \frac{M^2}{b^2},\\
&&\delta_{MqQ}=-\frac{3\pi \hat{q}Q\sqrt{1-v^2}}{v^2}\frac{M}{b^2},\\
&&\delta_{q^2Q^2}=\frac{\pi \hat{q}^2 Q^2 (1-v^2)}{2v^2}\frac{1}{b^2},\\
&&\delta_{aqQ}= \frac{2a\hat{q}Q\sqrt{1-v^2}}{v}\frac{1}{b^2}= \frac{a}{v}\frac{2qQ}{E}\frac{1}{b^2}.
\end{eqnarray}
Among these terms, $\delta_{M^2}$, $\delta_{MqQ}$ and $\delta_{q^2Q^2}$ are respectively the second order contributions from the gravitational interaction, gravitational-electrical coupling and pure electric interaction. They also appear in Reissner-Nordstr\"{o}m lensing of charged signals \cite{lidujia,Xu:2021rld}.

Here we point out that the importance of our result lies in the $\delta_{aqQ}$ term. It is known that the KN spacetime possesses a magnetic field asymptotically resembling the magnetic field caused by a dipole of moment $J=aQ$. On the equatorial plane it is given asymptotically by \cite{Misner:1974qy}
\begin{equation}
(B_r,B_\theta,B_\phi)=\left( 0,\frac{aQ}{r^3},0\right). \end{equation}
One can indeed show \cite{chen1977} that like $\delta_{aqQ}$, the deflection caused by this magnetic dipole to a relativistic charged particle in the equatorial plane in a flat spacetime is also proportional to $a\hat{q}Q\sqrt{1-v^2}/v$. Therefore we can identify $\delta_{aqQ}$ as the deflection caused by the magnetic dipole of the KN black hole.

The most interesting consequence of $\delta_{aqQ}$ becomes apparent when we compare the effect of $a$ on neutral and charged particles.
For neutral massive particle, $q=0$, from the term
\begin{align}
 \delta_{aM}=- \frac{4a}{v}\frac{M}{b^2}
\end{align}
of Eq. \eqref{asymptoticangle} it is seen that the spacetime spin $a$ would increases the deflection angle of retrograde particle ray, and decreases the deflection angle of prograde ray, as was known previously \cite{kejia}.
However, the effect of $a$ when $Q\neq0$ on the deflection of charged particles is different from that on neutral particles, due to the existence of the $\delta_{aqQ}$ term. Clearly, comparing $\delta_{aM}$ with $\delta_{aqQ}$, if $qQ/E<2M$, then the deflection angle of the charged signal is qualitatively still affected in the same way as neutral particles. However, if $qQ/E>2M$, the deflection angle would be increased for prograde particle ray, and decreased for retrograde particle ray. In particular, if $qQ/E=2M$, the terms $\delta_{aM^2}$ and $\delta_{aqQ}$ cancel thus $a$ does not contribute to the deflection angle at this order. Indeed, at this value of $qQ$, since
$\mathrm{sign}(qQ)=1$, the force between the lens and the signal is repulsive. Moreover, the value of $qQ$ is so large that when letting $qQ/E\rightarrow 2M$ in Eq. \eqref{asymptoticangle}, one can get
\begin{eqnarray}
\delta&=&\left( 1-\frac{1}{v^2} \right) \frac{2M}{b}+\frac{\pi}{4} \bigg[\left( 3-\frac{4}{v^2}\right)\nn\\
&&-\left( 1+\frac{2}{v^2} \right) \hat{Q}^2\bigg]\frac{M^2}{b^2}+ \mathcal{O}\left(\frac{[M]^3}{b^3}\right),
\end{eqnarray}
whose first order and therefore the entire $\delta$, obviously, is negative. This fact shows that the particle is divergently deflected by a KN black hole with such parameters $qQ/E\geq 2M$.

Furthermore, one can consider a rotating black hole with an electric charge $Q$ and a magnetic charge $P$, the so-called dyonic KN black hole, which has the same metric as the KN black hole with $Q^2$ replaced by $Q^2+P^2$~\cite{Kasuya1982}
\begin{eqnarray}
d s^{2} &=&\left(1-\frac{2 M r-(Q^2+P^2)}{\Sigma}\right) d t^{2}-\frac{\Sigma}{\Delta} d r^{2}-\Sigma d \theta^{2}\nn\\
&&-\sin ^{2} \theta d \phi^{2}\left[r^{2}+a^{2}+\frac{\left(2 M r-(Q^2+P^2)\right)}{\Sigma} a^{2} \sin ^{2} \theta\right] \nn\\
&&+2 a \sin ^{2} \theta \frac{2 M r-(Q^2+P^2)}{\Sigma} d \phi d t.\nn
\end{eqnarray}
where
\begin{eqnarray}
 &&\Sigma=r^{2}+a^{2} \cos ^{2} \theta,~~\Delta=r^{2}-2 M r+a^{2}+Q^{2}+P^{2}.\nn
\end{eqnarray}
The gauge field is given by
\begin{eqnarray}
A_{\mu} d x^{\mu}&=&\frac{Q r}{\Sigma}\left(d t-a \sin ^{2} \theta d \phi\right)\nn\\
&&+\frac{P}{\Sigma} \cos \theta\left[a d t-\left(r^{2}+a^{2}\right) d \phi\right].
\end{eqnarray}
In the equatorial plane ($\theta=\pi/2$), one finds that the part containing the magnetic charge $P$ vanishes in $A_{\mu}$. Therefore, the influence of magnetic charge $P$ on Jacobi geometry depends on the spacetime metric $g_{\mu\nu}$, but does not depend on the gauge field $A_{\mu}$ (see Eqs.\eqref{fensleranders1} and~\eqref{fensleranders2}). As a result, the deflection angle of charged particle by a dyonic KN black hole lens is
\begin{eqnarray}
\label{anglezpicture}
\delta&=&2\left( 1+\frac{1}{v^2} -\frac{\hat{q}\hat{Q}\sqrt{1-v^2}}{v^2}\right)\frac{M}{b}\nn\\
 &&+\bigg\{3\pi\left( \frac{1}{4}+\frac{1}{v^2} \right)-\pi\left( \frac{1}{4}+\frac{1}{2v^2}\right) \left(\hat{Q}^2+\hat{P}^2\right)\nn\\
 &&\mp\frac{4 \hat{a}}{v}-\frac{3\pi}{v^2}\hat{q}\hat{Q}\sqrt{1-v^2}\nn\pm\frac{2\hat{a}}{v}\hat{q}\hat{Q}\sqrt{1-v^2}\\
 &&+\frac{\pi}{2v^2} \hat{q}^2\hat{Q}^2\left(1-v^2\right) \bigg\}\frac{M^2}{b^2}+ \mathcal{O}\left(\frac{[M]^3}{b^3}\right).
\end{eqnarray}
There is no coupling between the magnetic charge and the charge of the particle. One can conclude that although the magnetic charge has an effect on the deflection angle, this effect makes no difference between neutral particle and charged particle. But it should be noted that the situation is different if one consider the deflection beyond the equatorial plane.

\section{Conclusion}\label{conclusion}

In this paper, we have explored the deflection angle of a charged particle by a KN black hole lens in the weak-field limit. The full second-order deflection angle of charged particle in KN spacetime is obtained in Eq. \eqref{asymptoticangle}, to our knowledge for the first time. It is revealed that to the leading order the spacetime spin $a$ manifests,  i.e., the $\mathcal{O}([M]^2/b^2)$ order, $a$ affects the deflection angle of charged particles both gravitationally through the $\delta_{aM}$ term and magnetically through the $\delta_{aqQ}$ term. The effect of $a$ on the deflection of charged particles is qualitatively different from that of neutral particle when $qQ/E>2M$: the deflection angle would be increased (or decreased) by $a$ for prograde (or retrograde) motion of the charged signal. If $qQ/E=2M$, parameters $a$ does not contribute to the deflection angle at the order $\mathcal{O}([M]^2/b^2)$ and the entire deflection is actually divergent due to the electric repulsion between the lens and the signal.
The dyonic KN black hole as a lens was also considered. The result shows that, on the equatorial plane, the magnetic charge $P$ has the same effect on the deflection of charged particles as on neutral particles.

To obtain the deflection angle, we used the Jacobi geometry for a charged massive particle in a stationary spacetime. The Jacobi geometry is defined by a Randers-Finsler metric, which has two other equivalent descriptions, i.e., Zermelo data and one dimension higher stationary spacetime data.
Because of the electromagnetic field, the motion of charged particle no longer follows a geodesic in the background spacetime. However, the trajectory of charged particle can be corresponded to geodesic in Randers space. Based on Randers data $(\alpha_{ij},\beta_i)$ and GB theorem, we used ORMM and the GJMM to obtain the deflection angle. It should be noted that in the latter method, the background space is defined by generalized Jacobi metric $\alpha_{ij}$, so the motion of the particles is non-geodetic. In addition, we calculated the deflection angle of null geodesics in $(n+1)$DSS using the iteration method, based on the fact that the geodesic in the $n$-dimensional Randers space can be regarded as the null geodesic in an $(n+1)$DSS~\cite{Gibbons2009}. In general, the two methods of using GB theorem link the geometric properties of a 3-space with the gravitational lensing, while the third method considers the null geodesic in $(3+1)$-spacetime. The results obtained by the three methods were shown to agree exactly.  There is also a fourth method, which uses the Hamilton-Jacobi equation to calculate the deflection angle in the background spacetime, see for example \cite{Jusufi:cmp} by Jusufi. In addition, the deflection of particles in the non-equatorial plane is particularly worth studying. We will leave this problem as the future project.

\acknowledgements

The authors thank Xiaoge Xv for help typesetting some of the formulas. This work is supported by the NNSF China 11504276 and MOST China 2014GB109004.

\appendix
\begin{widetext}
\section{The motion equation of charged particles in Kerr-Newman spacetime} \label{liuhaotian}
This appendix uses the perturbation method to solve from Eq. \eqref{hawking} the orbits of charged particles moving in the equatorial plane of KN spacetime. Equivalently, it can also be said to solve the orbits of photons in Jacobi 3-spacetime. For details of the method, readers can refer to Ref.~\cite{Arakida2012}. First, one can assume that in the large $b$ limit, the orbit takes a series form of $b$,
\begin{align}
\label{caironggen}
r(\phi)=r_1(\phi)b+r_0(\phi)+r_{-1}(\phi)b^{-1} \cdots
\end{align}
where $r_i(\phi)~(i=1,0,-1)$ are the coefficient functions of $\phi$ to be determined.
Then substituting this equation and the metric of the stationary spacetime \eqref{spacetime1} to \eqref{spacetime4} into Eq. \eqref{hawking}, carrying out the expansion in $b$ again and throwing away items of order two and higher in $1/b$, one can obtain ordinary differential equation for each $r_i(\phi)~(i=1,0,-1)$. The integral constants can be determined by taking the minimum value of $r$ at $\phi=\frac{\pi}{2}$, i,e., $\frac{dr_i}{d\phi}\big|_{\phi=\frac{\pi}{2}}=0$. Finally, the trajectory of the particle up to the second order in $1/b$ can be obtained with the coefficients
\begin{eqnarray}
\label{orbit}
&&r_1=\frac{1}{\sin \phi},~~~r_0=-\left(\cot^2 \phi+\frac{\csc^2 \phi}{v^{2}}\right)M+\frac{\sqrt{1-v^{2}} \csc^2 \phi}{v^{2}}\hat{q}Q ,\nn\\
&&r_{-1}=\frac{2 \csc^2 \phi}{v}Ma -\frac{\sqrt{1-v^{2}} \csc^3 \phi\left[6+\left(2+8 v^{2}\right) \cos (2 \phi)-3 v^{2}(\pi-2 \phi) \sin (2 \phi)\right]}{4  v^{4}}\hat{q}QM -\frac{\sqrt{1-v^{2}} \csc^2 \phi}{ v}\hat{q}Qa\nn\\
&&-\frac{\left[-4+16 v^{2}+2 v^{4} \cos^2 \phi+3 v^{2}\left(4+v^{2}\right)(\pi-2 \phi) \cot (\phi)-8\left(1+v^{2}\right)^{2} \cot^2 \phi\right] \csc \phi}{8 v^{4}}M^2\nn\\
&&+\frac{\left[4+2 v^{2} \cos^2 \phi+\left(2+v^{2}\right)(\pi-2 \phi) \cot\phi\right] \csc\phi}{8 v^{2}}Q^2+\frac{\left(1-v^{2}\right)\left(2\left(1-v^{2}\right)-v^{2}(\pi-2 \phi) \cot\phi+4 \cot^2\phi\right) \csc\phi}{4 v^{4}}\hat{q}^2Q^2.\nn\\
\label{eq:rnres}
\end{eqnarray}
According to this perturbation solution, we can also use radial coordinate to represent angular coordinate. Suppose the following formula
\begin{eqnarray}
\label{jiajunji}
\phi(r)=\begin{cases}
\phi^*(r),    &\text{if } \vert{\phi}\vert <\frac{\pi}{2};~~\\
\pi-\phi^*(r),&\text{if } \vert {\phi}\vert >\frac{\pi}{2}.~~
\end{cases} \label{eq:phitwosol}
\end{eqnarray}
and assuming that $\phi^*(r)$ takes the following quasi-series form of $b$,
\begin{eqnarray}
\label{anglecool}
\phi^*(r)&=&\phi_0+\phi_1\frac{M}{b}+\phi_2 \frac{\hat{q} Q}{b}+\phi_3 \frac{M a}{b^2}+\phi_4\frac{\hat{q} M Q}{b^2}+\phi_5 \frac{\hat{q} a Q}{b^2}+\phi_6 \frac{M^{2}}{b^2}+\phi_7 \frac{Q^{2}}{b^2}+\phi_8 \frac{\hat{q}^{2} Q^{2}}{b^2}+\phi_9\frac{a^2}{b^2}...,
\end{eqnarray}
we can substitute this equation into Eq. \eqref{caironggen} and solve iteratively $\phi_0$ to $\phi_9$. The results are
\begin{eqnarray}
\phi_0&=&\arcsin\left(\frac{b}{r}\right),~~~\phi_1=\frac{b^{2} v^{2}-r^{2}\left(1+v^{2}\right)}{r \sqrt{r^{2}-b^{2}} v^{2}},~~~\phi_2=\frac{r \sqrt{1-v^{2}}}{\sqrt{r^{2}-b^{2}} v^{2}}  ,~~~\phi_3=\frac{2 r}{\sqrt{r^{2}-b^{2}} v} ,\nn\\
\phi_4&=&-\frac{\sqrt{1-v^{2}}}{2 v^{4}}\left[\frac{2 b^{3}}{\left(r^{2}-b^{2}\right)^{3 / 2}}-\frac{6 b v^{2}}{\sqrt{r^{2}-b^{2}}}-3 v^{2}\left(\pi-2 \arcsin\left(\frac{b}{r}\right)\right)\right],~~~\phi_5=-\frac{r \sqrt{1-v^{2}}}{\sqrt{r^{2}-b^{2}} v},\nn\\
\phi_6&=&-\frac{3 b \sqrt{r^{2}-b^{2}}}{4 r^{2}}-\frac{3 b}{\sqrt{r^{2}-b^{2}} v^{2}}+\frac{b^{3}}{2\left(r^{2}-b^{2}\right)^{3 / 2} v^{4}}-\frac{3 v^{2}\left(4+v^{2}\right)\left(\pi-2 \arcsin\left(\frac{b}{r}\right)\right)}{8 v^{4}},\nn\\
\phi_7&=&\frac{b \sqrt{r^{2}-b^{2}}}{4 r^{2}}+\frac{b}{2 v^{2} \sqrt{r^{2}-b^{2}}}+\frac{\left(2+v^{2}\right)\left(\pi-2 \arcsin\left(\frac{b}{r}\right)\right)}{8 v^{2}},\nn\\
\phi_8&=&\frac{b^{3}\left(1-v^{2}\right)}{2\left(r^{2}-b^{2}\right)^{3 / 2} v^{4}}-\frac{b\left(1-v^{2}\right)}{2 \sqrt{r^{2}-b^{2}} v^{2}}-\frac{\left(1-v^{2}\right)\left(\pi-2 \arcsin \left(\frac{b}{r}\right)\right)}{4 v^{2}},~~
\phi_9=\frac{b^3}{2r^2\sqrt{r^2-b^2}}.
\end{eqnarray}

Finally, the deflection angle can be computed by taking the following limit
\begin{eqnarray}
\label{coordinateangle}
 \delta&=&-2\phi^*(r)\mid_{r\rightarrow \infty},
\end{eqnarray}
whose result is
\begin{eqnarray}
\label{eq:angleapp}
 \delta&=&2\left( 1+\frac{1}{v^2} -\frac{\hat{q}\hat{Q}\sqrt{1-v^2}}{v^2}\right)\frac{M}{b}+\bigg\{3\pi\left( \frac{1}{4}+\frac{1}{v^2} \right)-\pi\left( \frac{1}{4}+\frac{1}{2v^2}\right) \hat{Q}^2-\frac{4 \hat{a}}{v}\nn\\
 &&-\frac{3\pi}{v^2}\hat{q}\hat{Q}\sqrt{1-v^2}+\frac{2\hat{a}}{v}\hat{q}\hat{Q}\sqrt{1-v^2}+\frac{\pi}{2v^2} \hat{q}^2\hat{Q}^2\left(1-v^2\right)\bigg\}\frac{M^2}{b^2}+ \mathcal{O}\left(\frac{[M]^3}{b^3}\right).
\end{eqnarray}
Here $\hat{q}=q/m$ and $\hat{Q}=Q/M$ and $\hat{a}=a/M$.

\section{Components of Osculating Riemannian metric} \label{omga}
In this appendix, the components of the osculating Riemannian metric will be given. Making use of~\eqref{Hessian}, the Hessian of the Randers metric~\eqref{Randers-FinslerY} to \eqref{Randers-FinslerY3} can be obtained. Having found the Hessian, one can calculate the osculating Riemannian metric by substituting Eqs. \eqref{niubidun1} and~\eqref{niubidun2} into Eq. \eqref{lingling}. The result to the leading order(s) is found to be
\begin{eqnarray}
\label{wuhandaxue1}
\bar g_{rr} &=&\frac{q^{2} {Q}^{2}}{{r}^{2}}-\frac{2 {~m} {q} {Q}(4 {M}+{r})}{{r}^{2} \sqrt{1-{v}^{2}}}-\frac{{a} {m} {r} v\left(2 {~m} {M}-{q} {Q} \sqrt{1-{v}^{2}}\right) \sin^6\phi}{\left(1-{v}^{2}\right)\left({b}^{2} \cos^2 \phi+{r}^{2} \sin^4 \phi\right)^{\frac{3}{2}}}\nn\\
 &&-\frac{{m}^{2}\left[\left({a}^{2}-{r}^{2}\right) {v}^{2}+{Q}^{2}\left(1+{v}^{2}\right)-2 {Mr}\left(1+{v}^{2}\right)-4 {M}^{2}\left(2+{v}^{2}\right)\right]}{{r}^{2}\left(1-{v}^{2}\right)}+ \mathcal{O}\left(\frac{[M]^5}{[r]^3}\right),\\
\label{wuhandaxue2}
 \bar{g}_{r\phi}&=&\bar{g}_{\phi r}=\frac{a b^{3} m v\left(2 m M-q Q \sqrt{1-v^{2}}\right) \cos^3\phi}{r\left(1-v^{2}\right)\left(b^{2} \cos^2 \phi+r^{2} \sin^4\phi\right)^{\frac{3}{2}}}+ \mathcal{O}\left(\frac{[M]^2}{[r]^2}\right),\\
\label{wuhandaxue3}
\bar{g}_{\phi\phi}&=&{q}^{2} {Q}^{2}-\frac{2 {~m} {q} {Q}(2 {M}+{r})}{\sqrt{1-{v}^{2}}}+\frac{{m}^{2}\left(4 {M}^{2}-{Q}^{2}+2 {Mr}+\left({a}^{2}+{r}^{2}\right) {v}^{2}\right)}{1-{v}^{2}}\nn \\
 &&-\frac{{a} {m} {r} v\left(2 {~m} {M}-{q} {Q} \sqrt{1-{v}^{2}}\right) \sin^2 \phi\left(3 {~b}^{2} \cos^2 \phi+2 {r}^{2} \sin^4 \phi\right)}{\left(1-{v}^{2}\right)\left({b}^{2} \cos^2 \phi+{r}^{2} \sin^4 \phi\right)^{\frac{3}{2}}}+ \mathcal{O}\left(\frac{[M]^5}{[r]}\right).
\end{eqnarray}
Here in the higher order corrections we use $[r]^n$ to denote a combined order $n$ of $r$ and $b$. 

\section{The finite distance  deflection angle of charged particles in Kerr-Newman spacetime} \label{finidis}
In this appendix, we shall use the GJMM to compute the finite distance  gravitational deflection angle of charged particle by KN black hole. In this case, the distance $r_S$ from the particle source to the lens and the distance $r_R$ from the receiver to the lens are both finite. The angular coordinates of the source and receiver denoted as $\phi_S$ and $\phi_R$ respectively, satisfies the relation $\phi_R>\pi/2>\phi_S$. By Eq. \eqref{jiajunji}, we have
\begin{eqnarray}
\phi_S=\phi^*(r_S), ~~~~~~~\phi_R=\pi-\phi^*(r_R).
\end{eqnarray}
Replacing $D_\infty$ with a 2-space $\prescript{\infty}{R}\Box_{S}^{\infty}$ constructed for the finite distance source/receiver (see Refs.~\cite{OIA2017,Lijia2020} for details) in Eq. \eqref{kunlunjue}, the deflection angle can still be calculated by this formula,
\begin{eqnarray}
\delta=\delta_{gau}+\delta_{geo},
\end{eqnarray}
with only a modification of the integral limits of the following integration
\begin{eqnarray}
&&\displaystyle\delta_{gau}=-\iint_{\prescript{\infty}{R}\Box_{S}^{\infty}} K dS=-\int_{\phi_S}^{\phi_R} \int_{r(\phi)}^{\infty}K \sqrt{deta_{ij}} dr d\phi,\\
&&\displaystyle\delta_{geo}=\int_S^R \kappa(\eta) dl=\int_{\phi_S}^{\phi_R} \kappa(\eta) \frac{dl}{d\phi} d\phi.
\end{eqnarray}
Comparing with the calculation of the asymptotic deflection angle given by Eqs. \eqref{kunlunjue} to \eqref{eq:dgeores}, the integral limits here are  more general. However, the Gaussian curvature $K$ and geodesic curvature $\kappa(\eta)$ are  still given by  Eq. \eqref{gaussZermelo} and Eq. \eqref{piaoliuping}, respectively and the integration can still be carried out. Finally, the total finite distance deflection angle is found to be
\begin{eqnarray}
\label{qiaofeng}
 \delta&=&\delta_1\frac{M}{b}+\delta_2\frac{\hat{q}Q}{b}+\delta_3 \frac{\pi M^2}{b^2}+\delta_4 \frac{\ Q^2}{b^2}+\delta_5\frac{\hat{q}MQ}{b^2}+\delta_6\frac{ \hat{q}^2Q^2}{b^2}+\delta_7\frac{a\hat{q}Q}{b^2}+\delta_8\frac{Ma}{b^2}+ \mathcal{O}\left(\frac{[M]^3}{b^3}\right),
\end{eqnarray}
where
\begin{eqnarray}
 &&\delta_1=\left(1+\frac{1}{v^{2}}\right)\left(\sqrt{1-\frac{b^{2}}{{r_R}^{2}}}+\sqrt{1-\frac{b^{2}}{{r_S}^{2}}}\right),\nn\\
 &&\delta_2=-\frac{\sqrt{1-v^{2}}}{v^{2}}\left(\sqrt{1-\frac{b^{2}}{{r_R}^{2}}}+\sqrt{1-\frac{b^{2}}{{r_S}^{2}}}\right),\nn\\
 &&\delta_3=\frac{3}{4}\left(1+\frac{4}{v^{2}}\right)\left(\pi-\arcsin\left(\frac{b}{r_R}\right)-\arcsin\left(\frac{b}{r_S}\right)+\frac{b}{\sqrt{r_R^2-b^{2} }}+\frac{b}{\sqrt{r_S^2-b^{2}}}\right)\nn\\
 &&~~~~~~~-\left(\frac{3}{4}-\frac{1}{v^{4}}+\frac{2}{v^{2}}\right)\left(\frac{b^{3}}{r_R^{3} \sqrt{1-\left(\frac{b}{r_R}\right)^{2}}}+\frac{b^{3}}{r_S^{3} \sqrt{1-\left(\frac{b}{r_S}\right)^{2}}}\right),\nn\\
 &&\delta_4=-\left(\pi-\arcsin\left(\frac{b}{r_R}\right)-\arcsin\left(\frac{b}{r_S}\right)+\frac{b}{r_S} \sqrt{1-\frac{b^{2}}{r_S^2 }}+\frac{b}{r_R} \sqrt{1-\frac{b^{2}}{r_R^{2}}}\right)\left(\frac{1}{4}+\frac{1}{2 v^{2}}\right),\nn\\
 &&\delta_5=-\frac{\sqrt{1-v^{2}}}{v^2}\bigg[3\left(\pi-\arcsin\left(\frac{b}{r_R}\right)-\arcsin\left(\frac{b}{r_S}\right)\right)+2 b\left(1+\frac{1}{v^{2}}\right)\left(\frac{1}{\sqrt{r_R^2-b^2}}+\frac{1}{\sqrt{r_S^2-b^2}}\right)\nn\\
 &&~~~~~-\left(\frac{b^{3}}{r_R^{2} \sqrt{r_R^{2}-b^{2}}}+\frac{b^{3}}{r_S^{2} \sqrt{r_S^{2}-b^{2}}}\right)-\left(\frac{b}{r_S} \sqrt{1-\frac{b^{2}}{r_S^{2}}}+\frac{b}{r_R} \sqrt{1-\frac{b^{2}}{r_R^{2}}}\right)\left(\frac{2}{v^{2}}-1\right)\bigg],\nn\\
 &&\delta_6=\left(\frac{1}{2 v^{2}}-\frac{1}{2}\right)\left[\pi-\arcsin\left(\frac{b}{r_R}\right)-\arcsin\left(\frac{b}{r_S}\right)+\frac{2 b}{v^{2}}\left(\frac{1}{\sqrt{{r_R}^{2}-b^{2}}}+\frac{1}{\sqrt{r_S^{2}-b^{2}}}\right)\right]\nn\\
 &&~~~~~+\left(\frac{b}{r_S}\sqrt{1-\frac{b^{2}}{r_S^{2}}}+\frac{b}{r_R} \sqrt{1-\frac{b^2}{{r_R}^{2}}}\right)\left(1-\frac{2}{v^{2}}\right),\nn\\
 &&\delta_7=\frac{\sqrt{1-v^{2}}}{b^{2}  v}\left(\sqrt{1-\frac{b^{2}}{r_R^{2}}}+\sqrt{1-\frac{b^{2}}{r_S^{2}}}\right),\nn\\
 &&\delta_8=-\frac{2 }{v}\left(\sqrt{1-\frac{b^{2}}{r_R^{2}}}+\sqrt{1-\frac{b^{2}}{r_S^{2}}}\right).\nn
\end{eqnarray}
In the above, $\delta_1$ to $\delta_6$ are yielded by $\delta_{gau}$ and $\delta_7$ and $\delta_8$ are the results of $\delta_{geo}$. One can simply verify that taking the limits $r_R\rightarrow \infty$ and $r_S\rightarrow \infty$, Eq. \eqref{qiaofeng} reduces to the infinite distance deflection angle  Eq. \eqref{asymptoticangle}.
\end{widetext}

\end{document}